\documentclass[preprint2]{aastex}

\slugcomment{Accepted by Astrophysical Journal}

\shorttitle{Lifetimes of Gas and Dust Disks}
\shortauthors{Takeuchi, Clarke, \& Lin}

\begin{document}


\title{The Differential Lifetimes of Protostellar Gas and Dust Disks
\footnote{Accepted by ApJ}}

\author{Taku Takeuchi\altaffilmark{1,3}, C. J. Clarke\altaffilmark{2}, and
  D. N. C. Lin\altaffilmark{3}} 
\altaffiltext{1}{Earth and Planetary Sciences, Kobe University, Kobe 657-8501,
Japan; taku@kobe-u.ac.jp}
\altaffiltext{2}{Institute of Astronomy, Madingley Road, Cambridge CB3
  0HA, UK; cclarke@ast.cam.ac.uk}
\altaffiltext{3}{UCO/Lick Observatory, University of California, Santa Cruz,
CA95064; lin@ucolick.org}

\begin{abstract}

We construct a protostellar disk model that takes into account the
combined effect of viscous evolution, photoevaporation and the
differential radial motion of dust grains and gas.
For T Tauri disks, the lifetimes of dust disks that are mainly composed of
millimeter sized grains are always shorter than the gas disks'
lifetimes, and become similar only when the grains are
fluffy (density $\la 0.1 \ {\rm g \ cm}^{-3}$).
If grain growth during the classical T Tauri phase produces plenty of
millimeter sized grains, such grains completely accrete onto the star in
$10^7$ yr, before photoevaporation begins to drain the inner gas disk and
the star evolves to the weak line T Tauri phase.
In the weak line phase, only dust-poor gas disks
remain at large radii ($\ga 10$ AU), without strong signs of gas
accretion nor of millimeter thermal emission from the dust.
For Herbig AeBe stars, the strong photoevaporation clears the inner
disks in $10^6$ yr, before the dust grains in the outer disk migrate to
the inner region.
In this case, the grains left behind in the outer gas disk accumulate
at the disk inner edge (at $10-100$ AU from the star).
The dust grains remain there even after the entire gas disk has been
photoevaporated, and form a gas-poor dust ring similar to that observed
around HR 4796A.
Hence, depending on the strength of the stellar ionizing flux, our model
predicts opposite types of products around young stars.
For low mass stars with a low photoevaporation rate, dust-poor gas disks
with an inner hole would form, whereas for high mass stars with a high
photoevaporation rate, gas-poor dust rings would form. 
This prediction should be examined by observations of gas and
dust around weak line T Tauri stars and evolved Herbig AeBe stars.

\end{abstract}



\section{Introduction \label{sec:intro}}

Theoretical models for the evolution of protostellar disks have to satisfy
a number of observational constraints regarding the {\it relative}
timescales on which their dust and gas are dispersed. 
These constraints are:

 $(a)$ The majority of young stars of spectral type GKM exhibiting
evidence of {\it gas} accretion onto the stars (i.e., the classical T
Tauri stars, henceforth CTTSs; age $\sim 10^6$ yr) are observed to
possess dust disks of $\sim 100$ AU in size as evidenced by their
millimeter emission.
For example, Osterloh \& Beckwith (1995) detected 32 of the 62 CTTSs
($52 \%$) in their sample at a typical $3 \sigma$ level of $10-20$ mJy
(see also Beckwith et al. 1990). 
{\it Thus the dust in the outer disks cannot disperse below the present
detection limit on timescales much shorter than that of the gas in the
inner disks}.

$(b)$ By contrast, the majority of stars of spectral type GKM which lack
spectroscopic accretion indicators (i.e., the weak line T Tauri stars,
WTTSs)  are usually not detected at millimeter wavelengths [5 of 46
WTTSs ($11 \%$) in the sample of Osterloh \& Beckwith 1995; 2 of 10 WTTS
in Dutrey et al. 1996], even in recent high sensitivity 
surveys achieving $3 \sigma$ upper limits of a few mJy (1 of 12 in
Duvert et al. 1999).
{\it Thus the dust in the outer disks cannot disperse on timescales much
greater than that of the gas in the inner disks}.

At first sight, the simplest explanation of $(a)$ and $(b)$
is to invoke a constant dust-to-gas ratio both in space and in time, as
is usually assumed in modeling the disk evolution over $10^6$ yr
(e.g. Hartmann et al. 1998; Armitage et al. 1999, 2003; Clarke et
al. 2001).
If most grains are as small as $\sim 1 \ \micron$, the dust and the gas
are well coupled and so the constant dust-to-gas ratio over $10^6$ yr is
probably a reasonable assumption.
However, such an assumption is questionable for dust grains of
millimeter size or above, given their radial migration (Takeuchi \&
Artymowicz 2001; Takeuchi \& Lin 2002, 2003, and 2005, henceforth TL02,
TL03, and TL05, respectively).
Figure 1 of TL05 shows that the orbits of bodies in a size range of 1 mm
to 10 m decay in less than $10^6$ yr, which is the typical
age of CTTSs (Kenyon \& Hartmann 1995).
In addition, the constant dust-to-gas ratio does not appear to be
compatible with a third constraint, i.e., 

$(c)$ The transition timescale, over which stars lose circumstellar
diagnostics, is an order of magnitude less than the lifetime of the disk.
This conclusion is based on the relative paucity of `transition objects'
with infrared colors characteristic of optically thin inner disks
(Strom et al. 1989; Skrutskie et al. 1990; Kenyon \& Hartmann 1995).
The rapid turn-off of the inner disk, and of gas accretion diagnostics,
can be reproduced by models that combine viscous evolution with
photoevaporation by the central star (Clarke et al. 2001; Armitage et
al. 2003).
These models however predict a long lived outer gas disk (at radii $\ga 10$
AU). 
If the dust-to-gas ratio in the outer disk remains constant over the
typical age of CTTSs ($\sim 10^6$ yr), such evolution probably violates
constraint $(b)$, as discussed by Clarke et al. (2001).

In TL05, we showed that the long lived millimeter flux of CTTS
disks indicates that a sufficient amount of small grains less than $\sim 1$
mm must remain in the disks of $\sim 100$ AU scale for more than $10^6$ yr.
Otherwise, grains larger than 1 mm (and smaller than 10 m) rapidly
spiral inward to the star due to strong gas drag (see Fig. 1 of TL05). 
The population of the dust grains larger than 1 mm disappears quickly
from the disk, and thus the millimeter flux of the dust attenuates below the
observational detection limit within $10^6$ yr.
On the other hand, the evolution of the spectral index at millimeter
wavelengths suggests grain growth to at least millimeter sizes (Beckwith
\& Sargent 1991; Mannings \& Emerson 1994; Beckwith et al. 2000; Calvet
et al. 2002; Testi et al. 2003, however see also Agladze et al. 1996;
Mennella et al. 1998; Dupac et al. 2003 for other interpretations).
Combining these two considerations, we concluded that only a narrow range
of the grain size around 1 mm is responsible for the observed millimeter
fluxes of CTTSs.

Based on the above result, we further investigate the
relation of the lifetimes between the outer dust and inner gas disks.
In this paper, we show that, using plausible assumptions about the grain
population, the predictions of viscous photoevaporation models are
compatible with all the constraints $(a)$-$(c)$, once one accounts
for the differential radial motion of gas and dust predicted by simple grain
drag models. 
We describe such a `successful' model in which we assume (i) that the
dominant opacity at millimeter wavelengths derives from grains of sizes
close to $1$ mm and (ii) that there are no significant sources or sinks
of grains in that size range (i.e., that one may neglect the net
creation or destruction of such grains due to coagulation or the
collisional break up of grains).
We set out further observational predictions of our `successful'
model.


\section {A `Successful' Model for the Evolution of Dusty Disks}

\subsection {Evolution of the Gas}

We model the evolution of a gas disk due to the combined action of
viscosity and photoevaporation by ionizing radiation from the central
star.
We follow the models of TL05 and Clarke et al. (2001).
The initial gas density profile is $\Sigma_g = 3.5 \times 10^2(r/{\rm
AU})^{-1}  \ {\rm g \ cm}^{-2} $ with the outer radius of $r_{\rm out}=100$
AU.
The disk has a gas mass $M_g = 2.5 \times 10^{-2} \ M_{\sun}$ inside 100 AU.
The gas density obeys the equation,
\begin{equation}
\frac{\partial}{\partial t} \Sigma_g - \frac{3}{r}
\frac{\partial}{\partial r} \left[ r^{1/2} \frac{\partial}{\partial r} (
r^{1/2} \nu_t \Sigma_g) \right] = - \dot{\Sigma}_w \ .
\label{eq:cont_gas}
\end{equation}
The turbulent viscosity is expressed as $\nu_t=\alpha c^2/\Omega_{\rm K}
\propto r$, where $c$ is the sound speed and $\Omega_{\rm K}$ is the
Keplerian angular velocity. 
The temperature profile, $T= 278 \ (L/L_{\sun})^{1/4} (r / {\rm
AU})^{-1/2} \ {\rm K}$,  is assumed to be unchanged during the disk
evolution.
The sink term $\dot{\Sigma}_w$ in the right hand side is the wind
mass loss rate by photoevaporation, which has non-zero values 
outside the gravitational radius $r_g = G M/c_i^2 \sim 10^{14}$ cm,
where $M = 1 \ M_{\sun}$ is the central star's mass and $c_i=10^6 \ {\rm
cm \  s}^{-1}$ is the sound speed of the photo-ionized gas (Shu et
al. 1993; Hollenbach et al. 1994).
The wind mass loss rate per unit area is
\begin{equation}
\dot{\Sigma}_w = 2 c_i n_0 m_{\rm H} \cdot f\ ,
\end{equation}
where $n_0$ is the number density of hydrogen ions at the base of the 
photoevaporating flow, $m_{\rm H}$ is the mass of a hydrogen atom.
The base density is
\begin{equation}
n_0 = 5.7 \times 10^4 \Phi_{41}^{1/2} r_{g14}^{-3/2} \left(
\frac{r}{r_g} \right)^{-5/2} \ {\rm cm}^{-3} \ ,
\end{equation}
where $\Phi_{41}$ is the ionizing photon luminosity of the star in units
of $10^{41} \ {\rm s}^{-1}$, and $r_{g14}$ is the gravitational radius
in units of $10^{14}$ cm.
We introduce a function $f$ to express a drop in the wind mass loss rate
inside $r_g$,
\begin{equation}
f=\left[ 1+ \exp \left( - \frac{r-r_g}{\Delta r_g} \right) \right]^{-1}
\ , 
\end{equation}
where the transition width is assumed as $\Delta r_g = 0.1 r_g$.

In our model, the ionizing photon (EUV) flux $\Phi_{41}$ is constant with
time.
Because absorption by interstellar hydrogen atoms prevents us observing
EUV fluxes of young stars, there is no direct observational constraint
on it, and the origin of emitting EUV photons is also unclear.
If EUV photons are emitted from accretion shock at the stellar surface,
its flux decays with time and becomes inefficient as the disk accretion
ceases (Matsuyama et al. 2003; Ruden 2004).
Alexander et al. (2004) also showed that accretion shock
cannot emit enough EUV photons due to absorption by the stellar
atmosphere and by the accretion column.
Thus, chromospheric activity of young stars is probably responsible for
their EUV flux.
Recent analysis of the far-ultraviolet (FUV) line ratios in T Tauri
stars (Alexander et al. 2005) confirmed ionizing photon production rate
in the range $10^{41}-10^{43} \ {\rm s}^{-1}$.
In this paper, we simply assume a constant EUV photon luminosity
$\Phi_{41}$. 

\subsection {Evolution of the Dust}

The initial dust-to-gas ratio, $f_{\rm dust}$, is assumed to be constant
with $r$ and equal to $10^{-2}$.
Thus, the initial dust density profile is $\Sigma_d = 3.5 \times 10^2 f_{\rm
 dust}(r/{\rm AU})^{-1}  \ {\rm g \ cm}^{-2} $.
This is equivalent to assuming that the dust and gas mixture arrives in
the disk plane in the proportions in which they are found in the
 interstellar medium.
Naturally, we have no strong evidence that this is indeed the case and
that some fractionation of the dust and gas has not already occurred
during the collapse of the prestellar core and during the initial
evolution of the disk.
This issue however is beyond the scope of this paper.
We instead explore how the differential radial motion of dust and gas
alters the assumed initial constant dust-to-gas ratio.
The evolution of the dust surface density subject to
turbulent diffusion and gas drag is modeled according to TL05 as
\begin{equation}
\frac{\partial}{\partial t} \Sigma_d + \frac{1}{r}
\frac{\partial}{\partial r} \left[ r( F_{\rm dif} + \Sigma_d v_d )
  \right] = 0  \ ,
\label{eq:cont_dust}
\end{equation}
where $F_{\rm dif}$ is the diffusive mass flux caused by gas turbulence,
and $v_d$ is the dust radial velocity. 
We adopt the assumption that the diffusive mass flux is proportional to
the concentration gradient, and that the diffusion coefficient is the same
as the gas viscosity, which is applicable to small passive particles
tightly coupled to the gas, i.e., the stopping time of the grains due to
gas drag, $t_s$, is much smaller than the orbital time, $\Omega_{\rm
K}^{-1}$, and the Schmidt number ${\rm Sc}=1$.
In this case,
\begin{equation}
F_{\rm dif} = - \nu_t \Sigma_g \frac{\partial}{\partial r} \left(
  \frac{\Sigma_d}{\Sigma_g} \right) \ .
\end{equation}
The radial velocity of the dust subject to gas drag is calculated
according to equation (23) of TL02 as
\begin{equation}
v_d = \frac{T_s^{-1} v_g - \eta v_{\rm K}}{T_s + T_s^{-1}} 
\approx v_g - \eta T_s v_{\rm K}\ ,
\label{eq:dustvel1}
\end{equation}
where $v_{\rm K}$ is the Keplerian velocity, $T_s = t_s \Omega_{\rm
K}$ is the non-dimensional stopping time of the grain, $\eta$ is the
divergence factor of the gas rotation velocity from the Keplerian value.
The second equality is valid for $T_s \ll 1$.
The gas radial velocity $v_g$ is (eq. [11] of TL05)
\begin{equation}
v_g = - \frac{3}{r^{1/2} \Sigma_g} \frac{d}{dr} ( r^{1/2} \nu_t \Sigma_g)
\ .
\label{eq:gasvel}
\end{equation}
The non-dimensional stopping time of the grain $T_s$ that is
normalized by the local orbital time is (eq. [21] of TL05)
\begin{equation}
T_s = \frac{\pi \rho_p s}{2 \Sigma_g} \ ,
\label{eq:stoptime2}
\end{equation}
where $\rho_p$ is the grains' physical density and $s$ is the grain size.
The ratio of the radial gas pressure gradient to the gravity
is (eq. [20] of TL05 with $q=-1/2$)
\begin{equation}
\eta = -\frac{1}{r \Omega_{\rm K}^2 \rho_g} \frac{\partial P_g}{\partial
 r} =  -\left( \frac{h_g}{r} \right)^2 \left( \frac{r}{\Sigma_g} \frac{d
  \Sigma_g}{dr} - \frac{7}{4} \right) \ .
\label{eq:eta}
\end{equation}

In the above formulation we use midplane values and neglect the fact
that the gas 
rotation velocity and $\eta$ are functions of the height $z$ from the
midplane and that the stopping time $T_s$ also varies due to the gas
density change with $z$. 
This treatment is justified if grains sediment to the midplane quickly
enough.
From equation (14) of TL03, the timescale of sedimentation is
$\tau_{\rm sed} = |z/v_{z,d}| = (T_s \Omega_{\rm K})^{-1}$, where
$v_{z,d}$ is the vertical velocity of the grain.
This is smaller than the timescale of the grain's radial drift,
$\tau_{\rm dust} = |r/(v_d - v_g)| = \eta^{-1} (T_s \Omega_{\rm
K})^{-1}$, by a factor $\eta \ll 1$.
Sedimentation also proceeds more quickly than radial diffusion, whose
timescale is $\tau_{{\rm dif},r} = r^2/\nu_t \approx \alpha^{-1}
\eta^{-1} T_s (T_s \Omega_{\rm K})^{-1}$, if $ \alpha^{-1}
\eta^{-1} T_s > 1$.
In our standard model, $\alpha=10^{-3}$, $\eta \sim 10^{-2}$, and $T_s >
10^{-4}$ for millimeter sized grains at $r>10$ AU (see Fig. 2 of TL05),
and thus the condition $ \alpha^{-1} \eta^{-1} T_s \gg 1$ is satisfied.
Turbulent diffusion of moderate strength ($\alpha \sim 10^{-3}$) cannot
lift millimeter grains up beyond the disk scale height against their
sedimentation (for $r>10$ AU, see Fig. 4 of TL02).
 
\subsection{Opacity and Grain Size}

In this `successful' model, we consider a population of grains  with
size $1$ mm. 
The justification of this choice is that for a grain size distribution
$n(s) \propto s^{-\delta}$ with $\delta < 4$, the opacity at wavelength
$\lambda \sim 1$ mm is dominated by the largest grains, provided that
they are small enough to be described by the Rayleigh scattering theory
(i.e., $s \la \lambda$; see TL05).
Thus, if the grain size distribution extends up to $\sim
1$ mm in size, the grains can be modeled as a single sized
population of the largest grains as regards their opacity at millimeter
wavelengths. 

Evidently, this will not be a good approximation if the
maximum of the grain size distribution, $s_{max}$, is either
substantially less than or greater than $1$ mm. 
We note that the dust opacity index, $\beta$,
(which measures the frequency dependence of the dust
emissivity according to $\kappa_{\nu} \propto \nu^{\beta}$)
is $\la 1$ for CTTSs (Beckwith \& Sargent 1991; Mannings \& Emerson 1994;
Kitamura et al. 2002), which is consistent with grain
growth to at least $1$ mm. On the other hand, if the
dust were instead to reside mainly in much larger
grains, then the opacity at $1$ mm would be
considerably reduced (Fig. 4 in Miyake \& Nakagawa 1993) and could not
explain the observed millimeter fluxes from CTTSs.
 
Finally, in order to compute the disk emission at $1$ mm, we follow
Hartmann et al. (1998) and Clarke et al. (2001) in assuming that the
grain opacity is given by 
$\kappa_{\nu} = 0.1 (\nu / 10^{12} \ {\rm Hz} )$ and that
the dust temperature is a fixed power law function of the distance from
the central star [$T= 278 \ (L/L_{\sun})^{1/4} (r / 1 \ {\rm AU})^{-1/2}
\ {\rm K}$].


\section{Gas and Dust Disks' Lifetimes \label{sec:lifetime}}

\subsection {Model T Tauri Stars \label{sec:TTau}}

Following TL05 and Clarke et al. (2001), we take the parameters as $M=1 \
M_{\sun}$, $L=1 \ L_{\sun}$,  $r_{\rm out}=100$ AU,  $M_g = 2.5 \times
10^{-2} \ M_{\sun}$, $f_{\rm dust} = 10^{-2}$, $s = 1$ mm,
$\rho_p = 0.1-1 \ {\rm g \ cm}^{-3}$, $\alpha = 10^{-3}$, $r_g = 10^{14}$
cm, $\Phi_{41}=1$, and $c_i=10^6 \ {\rm cm \ s}^{-1}$, for T Tauri disks.
The equations for the gas and dust disks are solved in the same manner
as TL05.

The dashed lines of Figure \ref{fig:evo_ttau} depict the evolution of
the gas surface density  and illustrate that viscous evolution dominates the
emptying of the disk until the point is reached at which the accretion
rate becomes compatible with the mass loss rate due to photoevaporation
($\sim 10^{-10} \ M_{\sun}$ yr$^{-1}$ for an assumed ionizing luminosity
of $\Phi = 10^{41}$ s$^{-1}$). 
This mass loss, which is restricted to regions of the disk at radii $\ga
7$ AU (where photoionized material becomes unbound), is eventually
able to remove all of the mass flux from the outer disk and thus cuts
off the inner disk from re-supply. 
Consequently, the inner disk empties on the viscous timescale of the
disk at $\sim 7$ AU ($\sim 10^6$ yr for this model), which is much
less than the inner gas disk's life time of $\sim 10^7$ yr.
Gas however remains at radii $\ga 7$ AU for a further $10^7$ yr, this
being the timescale on which gas can viscously diffuse from the
outermost disk ($\sim 100$ AU) and reach radii of $\sim 10$ AU at which
photoevaporation becomes effective.

The solid lines of Figure \ref{fig:evo_ttau} depict the corresponding
evolution of the dust for two cases:
$(a)$ the grains' physical density, $\rho_p =1 \
{\rm g \ cm}^{-3}$ and $(b)$ $\rho_p = 0.1 \ {\rm g \ cm}^{-3}$.
Evidently, from equation (\ref{eq:stoptime2}), the lower density
`fluffy' grains are more tightly coupled to the gas.
Since for both grain densities, $T_s \ll 1$ (apart from where the gas
densities become very low in the inner disk at late times), equation
(\ref{eq:dustvel1}) shows that the fluffy grains will undergo slower
radial drift, being more tightly coupled to the radial motion of the
gas.
However, even for fluffy grains, the dust migration is significantly
faster than that of the gas for radii $\ga 10$ AU. 
This effect is much more marked for the compact grains, in which
case the dust evolves orders of magnitude faster than the gas. 
Figure \ref{fig:acc-flux} shows the evolution of the accretion rate of
the gas onto the star, plotted against the millimeter flux of the
thermal dust (calculated for face-on disks at a distance of 140 pc).
The dashed lines demonstrate that this
rapid evolution of 
compact grains violates observational constraint $(a)$ in
\S\ref{sec:intro}, regarding the high incidence of millimeter emission
in CTTSs.
The dust disks of compact grains evolve to very low millimeter fluxes
whilst the gas accretion rate is still high, thus generating a
population of CTTSs without millimeter emission. 
The slower migration of the fluffy grains however keeps the millimeter
flux above the detection limit ($\ga 10$ mJy) until the gas accretion
rate drops by a factor $3-10$.
The variations in the disk parameters such as the initial disk mass (the
initial surface density of the gas disk) and
the viscosity result in reasonable agreement with the observed
distribution of CTTSs in the plane of millimeter flux versus accretion
rate.

The dust evolution timescale is less than that of the gas for both types
of grains (we would need to reduce $\rho_p$ by a further order of
magnitude for the timescales to be comparable). 
This obviously means that constraint $(b)$ in \S\ref{sec:intro}
is easily satisfied.
Although the disk photoevaporation models predict that gas resides at
large radii ($\ga 10$ AU) for $\sim 10^7$ yr after the cessation of
accretion onto the star, such gas is, in these models, heavily depleted
in dust.
The predicted millimeter flux from the outer disk is orders of
magnitude less than the current detection limit at the stage that the inner
disk is cleared out.

In summary, we find that models that combine viscous evolution,
photoevaporation and gas-dust drag will readily satisfy constraints
$(b)$ and $(c)$ in \S\ref{sec:intro}, but that constraint $(a)$ is only
satisfied in the case of fluffy grains (density  $\rho_p \la 0.1$ g
cm$^{-3}$).
 
\subsection{Model Herbig AeBe Stars}

In this model, the central star is $M=2.5 \ M_{\sun}$ and $L=30 \ L_{\sun}$
(a typical value for Herbig AeBe stars; van den Ancker et al. 1998).
The gravitational radius, $r_g$, beyond which ionized gas is unbound,
increases by a factor $2.5$. 
The initial value of the gas surface density is the same as the
T Tauri star model, but the disk outer radius is increased to $r_{\rm
out}=250$ AU.
The total disk mass is $6.25 \times 10^{-2} \ M_{\sun}$.
Most significant, however, is the increase in ionizing luminosity.
The observation of FUV ($h \nu < 13.6 \ {\rm eV}$) from young stars by
Valenti et al. (2003) showed that continuum fluxes at
$\lambda = 2257$ \AA \ from Herbig AeBe stars are about two orders of
magnitude larger than those of CTTSs on average.
A stellar wind model by Bouret \& Catala (1998) suggests that the ionizing
fluxes of Herbig AeBe stars are of order of $\Phi_{41}=10^2-10^4$.
In our model, we assume $\Phi_{41}=10^3$.

The net photoevaporation mass loss rate, which is proportional to $\sqrt{r_g
\Phi_{41}}$ (Hollenbach et al. 1994), is increased by a factor
$\sqrt{2500}$ with respect to the T Tauri star model.
The model thus evolves to the point that the inner disk is cleared out within
a couple of Myr, an order of magnitude faster than the T Tauri star case. 
Consequently, the disk is much less depleted in the dust at the point that
the inner disk is drained away.
Figure \ref{fig:evo_haebe} demonstrates there is a substantial column in
the dust in the outer disk at this stage.
Evidently, once the wind truncates the disk at its inner edge (initially
close to $r_g \approx 17$ AU), the dust can no longer migrate toward the
star and instead accumulates in a dust ring close to the inner edge of
the gas disk.
We note that as the wind progressively eats away at the
outer disk, the inner edge of the gas disk grows to radii $r \gg r_g$ (as
in the models of Clarke et al. 2001) and that this is accompanied by the
dust distribution becoming increasingly annular, as the dust is concentrated
by inward migration, near the inner edge of the gas disk.
The steep increase in the gas density and pressure with $r$ near the
inner edge induces a super-Keplerian velocity of the gas, and
makes the inward velocity of the dust reverse in this region. 
This phenomena of dust accumulation is similar to those discussed by
Klahr \& Lin (2001) and by Haghighipour \& Boss (2003a,b).
It is intriguing to note the similarity of this progression
to the observed dust distributions in Herbig AeBe stars,
such as the disk with a small inner hole in HD 100546 ($\sim 10$ AU;
Bouwman et al. 2003) and the dust ring at a larger radius in HD 141569A
(the inner edge $\sim 175$ AU; Weinberger et al. 1999; Augereau et
al. 1999a; Clampin et al. 2003).
 
We thus find that through changing the parameters of the central star,
but leaving the grain model unchanged, we can simultaneously account for the
absence of long lived millimeter emission in WTTSs [constraint $(b)$ in
\S\ref{sec:intro}] whilst predicting outer dust disks of Herbig AeBe stars
that have ceased to accrete.
The essential difference is that the higher EUV flux of Herbig AeBe
stars is able to significantly affect the disk on a much shorter
timescale, so that the dust has not had time to migrate onto the star
at the point that the inner disk becomes cleared.

\subsection{Other Parameters \label{sec:para}}

We now briefly consider how the above results depend on model
parameters, while still retaining the assumption of a single sized
grain population that is subject to neither grain creation nor destruction 
processes.
We find that we can vary the lifetime of the disk through varying a
number of parameters (for example, the disk initial mass, radius, or
$\alpha$ viscosity parameter).
The properties of the grains enter the evolution equations only via
the combination $\rho_p s$, which determines the degree of coupling
between the grains and the gas (i.e., the stopping time in
eq. [\ref{eq:stoptime2}]). 
We might thus expect that the ratio of the evolution timescales of the  
dust and the gas should depend on this product, and therefore
define a parameter
\begin{equation}
A \equiv 
\left( \frac{s}{1 \ {\rm mm}} \right)
\left( \frac{\rho_p}{1 \ {\rm g \ cm}^{-3}} \right) \ .
\label{eq:A1}
\end{equation}

This expectation is confirmed by Figure \ref{fig:ltime}, which plots the
outer dust disk lifetime against the inner gas disk lifetime for a range
of disk models. 
The standard parameters of the model are the ones described in
\S\ref{sec:TTau} for T Tauri disks, and the disk mass, the outer
radius, the viscosity, and the dust parameter $A$ are varied.
Following Armitage et al. (2003), we use the gas accretion
rate to determine the inner gas disk lifetime.
The presence of gas accretion is generally considered as an observable
indicator of having a gas disk.
In our model, the gas lifetime is defined as the time at which the gas
accretion onto the star drops to less than $10^{-10} \ M_{\sun} \ {\rm
yr}^{-1}$.
As shown by Clarke et al. (2001), when the inner disk is cleared out,
gas accretion vanishes.
Hence, the gas lifetime defined here corresponds to the inner disk lifetime,
although the outer disk might remain much longer.
On the other hand, the outer dust disk lifetime is defined by the luminosity
of $1.3$ mm continuum.
When the luminosity $4 \pi D^2 \nu F_\nu$ drops below $10^{28} \ {\rm erg
  \ s}^{-1}$, the dust disk is defined to be removed.
This criterion approximately corresponds to the detection limit of $2.5$
mJy at $D=140$ pc, which is the limit of the observation by Duvert et al.
(2000).

Clearly, the results lie in three bands of dust lifetime/gas lifetime,
differentiated by the value of $A$. 
We thus see that, for the single sized grains of $\sim 1$ mm,
although the {\it absolute} lifetimes are dependent on the disk
parameters such as the viscosity, the initial mass, and the radius,
the {\it relative} lifetimes are just functions of $A$.
Thus the fractional incidence of millimeter emission amongst 
CTTSs is controlled by $A$.
This means the conclusion that millimeter sized grains must be 
rather fluffy ($\rho_p \la 0.1$ g cm$^{-3}$) in order to satisfy constraint
$(a)$ in \S\ref{sec:intro} is a rather general one.

\subsection{Changing the Grain Size \label{sec:size}}

So far we have considered the case of grains of size 1 mm that
are subject to neither replenishment nor destruction by coagulation or
shattering.
In CTTSs, the small values of opacity index at millimeter wavelengths,
$\beta \la 1$, (where $\kappa_\nu \propto \nu^{\beta}$), imply that the
grain size distribution extends to at least millimeter scales
(Miyake \& Nakagawa 1993; see however, e.g.,  Beckwith et al. 2000 for
other interpretations).
Thus, we do not further consider the case that the size $s \ll 1 $ mm. 
On the other hand, if the grain size is too large ($s \gg 1$ mm), the
opacity at millimeter wavelengths is reduced considerably (Miyake \&
Nakagawa 1993).

If we now instead consider the case that $s \sim 1$ cm, we have
a problem, as discussed in detail by TL05. 
These grains undergo 10 times faster radial migration than millimeter
sized grains and cannot survive over $10^6$ yr.
If the opacity at millimeter wavelengths relies on the presence of centimeter
sized grains, then, in the absence of replenishment, the timescale
over which CTTSs would exhibit millimeter emission would be very short
(less than $10^6$ yr), in violation of constraint $(a)$.
Thus, in the absence of grain replenishment, we argue against the
case $s \sim 1$ cm.


\section{Discussion}

\subsection{Dust Grains Loss by the Photoevaporating Wind} 

The photoevaporating wind may carry the dust grains away from the
disk.
The gas drag force of the wind on a grain is $F \approx \case{4}{3} \pi
n_0 m_{\rm H} s^2 c_i^2 = m_d n_0 m_{\rm H} c_i^2 / (\rho_p s)$, where
$m_d$ is the spherical grain's mass.
For the Herbig AeBe star case, where $M=2.5 \ M_{\sun}$, $r_g=2.5 \times
10^{14}$ cm, $\Phi_{41}=10^3$, and $\rho_p = 0.1 \ {\rm g \ cm}^{-3}$, the
drag force is weaker than the gravity $G M m_d/r_g^2$, as long as $s \ga
20 \ \micron$.
Hence, if the dust grains have grown larger than $20 \ \micron$, most
of the dust mass remains in the disk even after the gas disk has completely
evaporated.
In this case, it is expected that a ring-like dust disk remains after
the gas dispersal, like the HR 4796A dust-ring (Schneider et al. 1999).
Further evolution of the dust ring during or after the gas dispersal
likely alters its morphology, for example, by radiation pressure
(Klahr \& Lin 2001; Takeuchi \& Artymowicz 2001), by collisions of
grains (Kenyon \& Bromley 2004), and by unseen planets (Wyatt et
al. 1999).
(See also Augereau et al. 1999b; Li \& Lunine 2003; Currie et al. 2003;
Chen \& Kamp 2004, for modeling of the HR 4796A disk based on its
emission.)
Detailed discussion on such later evolution stages are, however, beyond
the scope of this paper.

For T Tauri disks of $\Phi_{41}=1$, the drag force by the wind is much
weaker, and thus even $1 \ \micron$ grains can resist the wind.
Hence, grain loss by the inward migration is much more important in
T Tauri disks.

\subsection{The Possible Role of Grain Replenishment} 

We concluded in \S\ref{sec:para} and \S\ref{sec:size} that the bulk of
the opacity at millimeter wavelengths is provided by grains of sizes
similar to or less than about 1 mm in order to satisfy constraint $(a)$.
Regarding the fraction of CTTSs with detectable millimeter emission, we
need to retain a stock of small ($s \la 1$ mm) grains in the disk over
several Myr.  
We have shown in \S\ref{sec:lifetime} that such grains could avoid
excessively rapid migration if they were rather fluffy, but there is also
a possibility that they are otherwise depleted by coagulation to larger
grain sizes.
This possibility was considered in TL05.
In this subsection, we note that this problem would be circumvented if the
stock of grains of $s \la 1$ mm were replenished, either by continued infall
of small grains from a circumstellar envelope or by shattering of
larger bodies.     

We cannot rule out that this is the case.
However, we note that although this would certainly satisfy constraint
$(a)$, it would, without fine tuning, fail to satisfy constraint $(b)$.
In other words, if opacity at millimeter wavelengths were sustained in this
way throughout the CTTS phase, it would over produce the millimeter
emission from the outer disk in the WTTS phase. The only way
that this could be avoided would be if, coincidentally, the
transition from the CTTS phase to the WTTS phase also turned
off the replenishment process. It is not clear, however, why this should
be the case.


\section{Conclusions}

We have shown that the disk model that combines viscous evolution,
photoevaporation and the differential radial motion of grains and gas can
account for the three observational constraints $(a)$-$(c)$ set out in
\S\ref{sec:intro}.
We particularly highlight the fact that the same model of the grains can
simultaneously account for the fact that the non-accreting T Tauri stars
(WTTSs) do not possess detectable millimeter emission, whereas some
of their non-accreting counterparts in the Herbig AeBe mass range
show prominent dust disks at large radii.

The difference between the two types of behavior is due to the fact
that Herbig AeBe stars suffer much stronger photoevaporative mass loss
from the disk (due to the stronger EUV flux) and this causes accretion
onto the star to cease at a much younger age. 
This relatively rapid draining of the inner gas disk means that
there is insufficient time for the grains to have migrated onto the
star.
Consequently, the grains are stranded in the outer disk (i.e., at 
$r > r_g$).
As the outer disk is slowly eroded by photoevaporative mass loss,
the inner hole in the disk grows, while at the same time the dust
becomes increasingly concentrated toward the inner edge of the residual
gas disk.
Thus the appearance of dust rings at large radii (tens of AU),
which is sometimes attributed to the action of planets,
is a natural consequence of photoevaporation and grain migration.
In T Tauri stars, by contrast, the photoevaporative flow is much
gentler.
By the time this flow has created an inner hole in the gas disk, the dust has
all migrated onto the star, and so WTTSs are not generally millimeter
wavelength sources.
It is intriguing in this regard that the small number of T Tauri stars
with optically thin inner disks which yet show millimeter emission
from the disk at large radii are objects with unusually high ultraviolet
fluxes (Bergin et al. 2004).
This supports our conjecture that the timescale for mass loss due to
photoevaporation, relative to the timescale for grain migration, is an
important factor in the formation of these systems.

We find that the hardest of the observational constraints to satisfy is
that the millimeter flux in T Tauri stars should 
persist for a good fraction of the time that the star
spends as a CTTS.
This places a constraint on the relative timescales for dust and gas
evolution, which, in our models, is determined almost entirely by the
product of the grain density and the grain size, $\rho_p s$, since this
controls the degree to which the grain motions are coupled to the gas. 
We find that in order for the millimeter flux to persist long enough we
need to invoke a population of grains that is rather fluffy ($\rho_p \la
0.1$ g cm$^{-3}$).

\begin{figure}
\epsscale{1.0}
\plotone{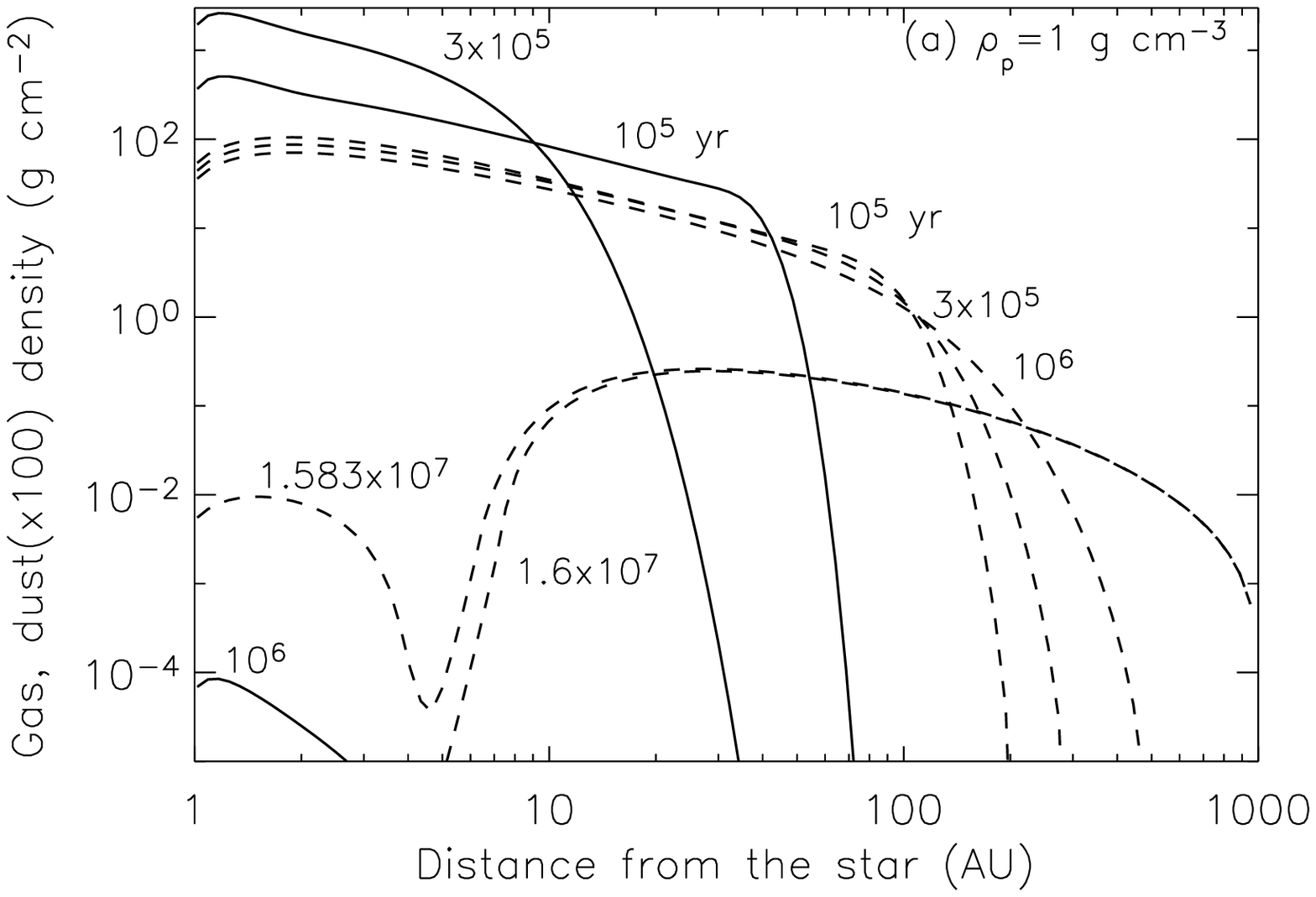}
\plotone{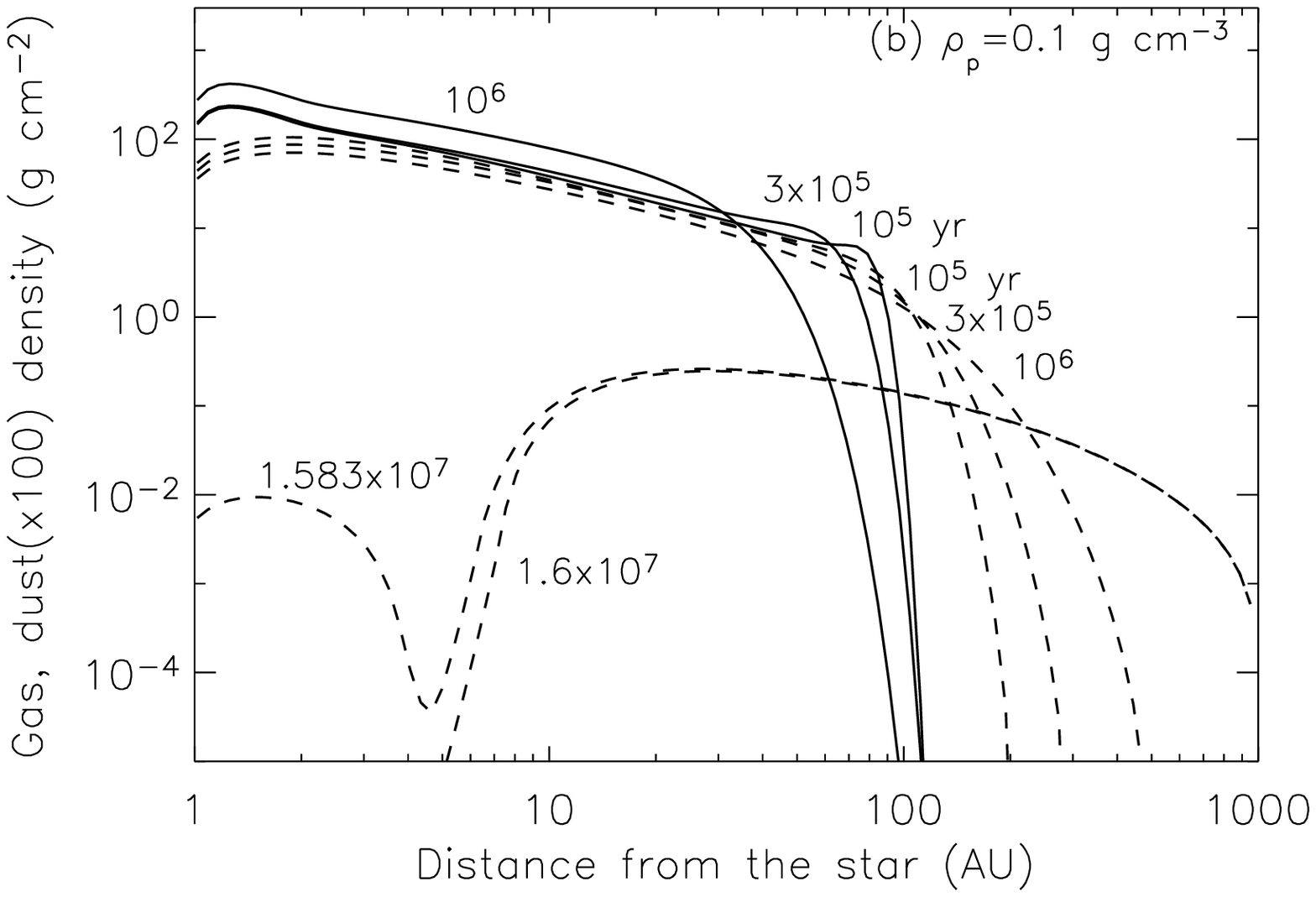}
\caption{
Evolution of the gas and dust surface densities of a T Tauri disk.
The dust surface densities are multiplied by the gas-to-dust ratio
($f_{\rm dust}^{-1}=100$) and shown by the solid lines.
The dashed lines show the gas surface densities.
 ($a$) Compact grains, $\rho_p = 1 \ {\rm g \ cm}^{-3}$. 
Most of the dust grains have disappeared by $10^6$ yr.
($b$) Fluffy grains, $\rho_p = 0.1 \ {\rm g \ cm}^{-3}$.
The dust grains remain over $10^6$ yr, but they cannot survive until
photoevaporation begins to clear the inner gas disk at $1.6 \times 10^7$ yr.
\label{fig:evo_ttau}
}
\end{figure}

\begin{figure}
\epsscale{1.0}
\plotone{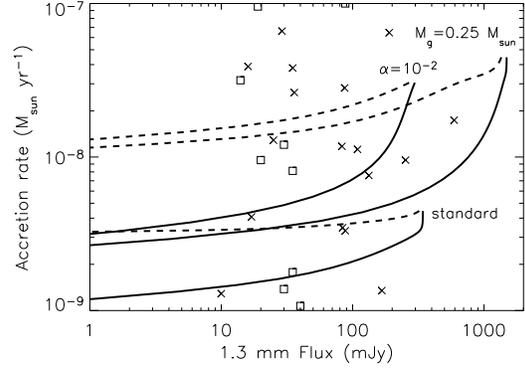}
\caption{
Gas accretion rate vs. dust thermal flux at a wavelength of $1.3$ mm.
The solid lines are for disks with fluffy grains ($\rho_p
= 0.1 \ {\rm g \ cm}^{-3}$), while 
the dashed lines are for compact grains ($\rho_p = 1 \ {\rm g \ cm}^{-3}$).
The evolution of three models [the standard model, the massive disk
model of $M_g=0.25 \ M_{\sun}$ (10 fold increase in the initial gas
density), and the high viscosity model of $\alpha =10^{-2}$] is plotted.
The $1.3$ mm fluxes are calculated for face-on disks at a distance  of
$D=140$ pc.
Crosses are the observed values of T Tauri stars.
The 1.3 mm fluxes are taken from Beckwith et al. (1990) and
Osterloh \& Beckwith (1995), and the gas accretion rates are taken from
Hartmann et al. (1998).
Squares show the upper limits of $1.3$ mm fluxes.
\label{fig:acc-flux}
}
\end{figure}

\begin{figure}
\epsscale{1.0}
\plotone{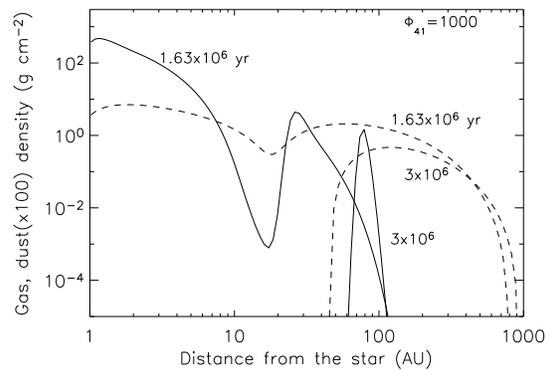}
\caption{
Same as Fig. \ref{fig:evo_ttau}, but for a Herbig AeBe disk.
The grains are fluffy ($\rho_p = 0.1 \ {\rm g \ cm}^{-3}$), and
the ionizing photon luminosity is $\Phi_{41}=10^3$.
\label{fig:evo_haebe}
}
\end{figure}

\begin{figure}
\epsscale{1.0}
\plotone{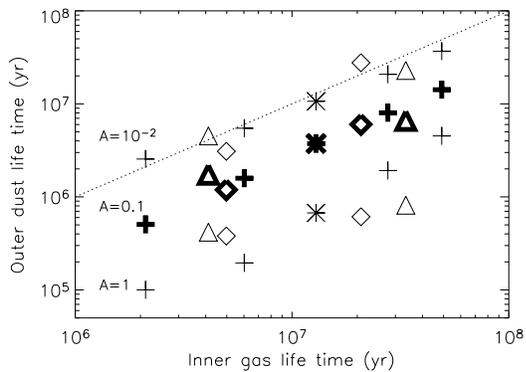}
\caption{
Lifetimes of the gas and dust disks of T Tauri star models.
The upper group of the thin symbols are for $A=10^{-2}$, the thick
symbols are for $A=0.1$, and the lower thin symbols are for $A=1$. 
Asterisks ($\ast$) are for the standard gas disk with $M_g=2.5 \times
10^{-2} \ M_{\sun}$, $r_{\rm out}=100$ AU, and $\alpha=10^{-3}$. 
Crosses (+) are for disk masses, $M_g=2.5 \times 10^{-3}$, $8 \times
10^{-3}$, $8 \times 10^{-2}$, and $2.5 \times 10^{-1} \ M_{\sun}$ from the
left. 
Triangles ($\triangle$) are for disk radii, $r_{\rm out}= 30$ AU and 300
AU from the left.
Diamonds ($\Diamond$) are for viscosities, $\alpha=10^{-2}$ and
$10^{-4}$ from the left.
The dotted line represents that the lifetimes of the gas and the dust
are the same.
\label{fig:ltime}
}
\end{figure}

Finally, we note a clear observational prediction of the type of model
that we have described here.
In these models, WTTSs are systems where the inner disk has drained onto
the star, having been starved of resupply by photoevaporative mass loss
from the outer disk.
WTTSs then retain disk gas at large radii for considerable periods
(Clarke et al. 2001), but we argue here that the dust content of such
disks should be low, due to the prior radial migration of the dust. 
In this way, we circumvent the problem that WTTSs show no detectable
dust emission in general. 
We however predict that these systems should retain a significant column
density in {\it gas} during the WTTS phase (see Fig. \ref{fig:evo_ttau}).
We point out that detection of gas diagnostics in WTTSs at large radii
would provide striking confirmation of this type of model.

\acknowledgements
We wish to thank Yoichi Itoh, Ted Bergin, and Richard Alexander for
useful discussions and Anthony Toigo for helpful comments.
We acknowledge helpful comments from the referee.
This work was supported in part by an NSF grant AST 99 87417 and in part
by a special NASA astrophysical theory program that supports a joint
Center for Star Formation Studies at UC Berkeley, NASA-Ames Research
Center, and UC Santa Cruz.
This work was also supported by NASA NAG5-10612 through its Origin
program and by JPL 1228184 through its SIM program, and by
the 21st Century COE Program of MEXT of Japan, ``The Origin and
Evolution of Planetary Systems.'' 
CJC is grateful for the hospitality of
UC Santa Cruz during the initial stages of this work and
acknowledges support from the Leverhulme Trust in the form
of a Philip Leverhulme Prize.



\clearpage
 
\end{document}